\documentstyle[12pt]{article}
\begin{document}
\begin{titlepage}
  \begin{flushright}
    KUNS-1700\\[-1mm]
  \end{flushright}
  \vspace{5mm}
  
  \begin{center}
    {\large\bf Can we distinguish SUSY breaking terms between \\
weakly and strongly 
coupled heterotic string theories ? \\
}
    \vspace{1cm}
    
    Tatsuo~Kobayashi\footnote{E-mail address:
      kobayash@gauge.scphys.kyoto-u.ac.jp} 
    \vspace{5mm}
    
    {\it Department of Physics, Kyoto University
      Kyoto 606-8502, Japan}
    \vspace{1.5cm}
    
    \begin{abstract}
We study soft SUSY breaking terms in heterotic M-theory.
We show that both weakly and strongly coupled 
heterotic string models lead to the same relations of 
soft SUSY breaking terms, 
$A=-M$ and $m^2 = M^2/3$, up to $O((\alpha T/S)^2)$.
    \end{abstract}
  \end{center}
\end{titlepage}

\section{Introduction}

Supersymmetry (SUSY) breaking is important to 
connect superstring theory with our real world.
Under the assumption that the dilation field $S$ and 
moduli fields $T^i$ contribute to SUSY breaking, 
soft SUSY breaking terms have been 
studied within the framework of effective supergravity theory 
inspired by superstring theory, in particular weakly coupled heterotic 
string theory \cite{CCM}-\cite{BIMS}.
Their phenomenological implications have been investigated from 
various viewpoints.
The weakly coupled heterotic string theory leads to 
the following relations 
among soft SUSY breaking terms, 
\begin{equation}
h^{ijk} = - M\hat Y^{ijk}, 
\label{M-A}
\end{equation}
\begin{equation}
m_i^2 + m_j^2+m_k^2 =|M|^2,
\label{M-m}
\end{equation}
where $M$ is the gaugino mass, $h^{ijk}$ is the trilinear 
coupling of scalar fields corresponding to the Yukawa coupling 
$\hat Y^{ijk}$ and 
$m_i$ is the soft scalar mass of scalar field $\phi_i$
for non-vanishing Yukawa couplings $\hat Y^{ijk}$
if Yukawa couplings in the supergravity basis 
are field-independent, the scalar potential 
is $T$-duality invariant and the vacuum energy vanishes.
In particular, the case with universal K\"ahler metric 
of chiral matter fields, e.g. orbifold models with only 
untwisted sectors and the large $T$ limit of Calabi-Yau models,  
leads to the universal scalar mass, $m_0$,
\begin{equation}
m_0^2 = {1 \over 3}|M|^2.
\label{M-m2}
\end{equation}

These relations (\ref{M-A}) and (\ref{M-m}) are one-loop
renormalization group (RG) 
invariant if the Yukawa coupling is on the 
RG-invariant trajectory $\hat Y^{ijk} = g C^{ijk}$, 
where $g$ is the gauge coupling and $C^{ijk}$ is a constant 
determined by group-theoretical factors \cite{JJ,sum-rule}.
$T$-duality anomaly induces a certain correction to 
the relations.
On the other hand, the two-loop and all order RG-invariant relations 
have been derived in Refs.\cite{kkmz,kkz}, and those are exactly 
coincide with the relations induced by the $T$-duality anomaly 
for finite theories.
These results on the soft SUSY breaking terms 
might be a remnant of finiteness of string theory.

Similarly, effective supergravity theory inspired by 
strongly coupled heterotic theory, M-theory on $S^1/Z_2$ \cite{M-theory}, 
has been obtained \cite{banks}-\cite{ELP}.
That is different from the weakly coupled one by terms including 
the factor $\alpha T/S$.
In addition, soft SUSY breaking terms have been discussed 
with use of this effective supergravity 
theory \cite{nilles2}-\cite{li3}.\footnote{For phenomenological
aspects, see e.g. Ref.\cite{BKL}.}
The results seem to differ from those of weakly coupled theory 
by the factor, $\alpha T/S$.
Thus, in the literature it has been deduced that the heterotic M-theory would 
lead to phenomenological 
aspects different from the weakly coupled one.

In this paper we study soft SUSY breaking terms 
derived from heterotic M-theory and we will show 
heterotic M-theory also leads to 
the relations (\ref{M-A}), (\ref{M-m2}) up to $O((\alpha T/S)^2)$.
Thus, at this level we can not distinguish 
soft SUSY breaking terms between 
weakly and strongly coupled heterotic string models.

\section{Weakly coupled heterotic models}

At first we give a brief review on soft SUSY breaking terms 
derived from generic supergravity theory.
We start with the following K\"ahler potential $K$ and 
the superpotential $W$,
\begin{eqnarray}
K &=& \tilde K(\Phi_a,\bar \Phi^a)+ K^i_i(\Phi_a,\bar \Phi^a)
|\Phi_i|^2, \\
W &=& \tilde W(\Phi_a)+ {1 \over 6}Y^{ijk}\Phi_i\Phi_j\Phi_k,
\end{eqnarray}
where chiral superfields have been classified 
into two types: $\Phi_a$ would have large vacuum expectation 
values and the other light fields are denoted by $\Phi_i$. 
All of them are denoted by $\Phi_I$.
Here $\tilde W(\Phi_a)$ is a nonperturbative superpotential 
leading to SUSY breaking.
We have neglected supersymmetric mass terms of $\Phi_i$.
The scalar potential is obtained as 
\begin{equation}
V = F_I F^J K^I_J - 3 e^G,
\end{equation}
where $F_I$ are $F$-terms of $\Phi_I$ and $G=K+\ln|W|^2$.
Here we assume that the fields $\Phi_a$ develop their vacuum 
expectation values and their 
$F$-terms contribute to SUSY breaking.
Then, taking the flat limit, 
we can expand the scalar potential around nonvanishing 
$\Phi_a$ and $F_a$,
\begin{equation}
V=V_0 + \sum_i\left|{\partial W_{eff} \over \partial \phi_i}\right|^2 + 
{1 \over 2}m^2_i|\phi_i|^2 K^i_i+({1 \over 6} h^{ijk}\phi_i \phi_j \phi_k 
\prod_{\ell=i,j,k}(K^\ell_\ell)^{1/2} + h.c.) + \cdots,
\label{s-V}
\end{equation}
where $\phi_i$ is the scalar component of $\Phi_i$, and 
the first term of RHS is the vacuum energy obtained as 
$V_0= F_a F^b K^a_b - 3 m_{3/2}^2$ with $m_{3/2}^2 \equiv e^G$.
The second term of RHS is globally supersymmetric terms where 
the Yukawa coupling $\hat Y^{ijk}$ in the global SUSY basis 
is obtained for the canonical field, 
\begin{equation}
|\hat Y^{ijk}|^2 = e^{\tilde K}|Y^{ijk}|^2 (K^i_iK^j_jK^k_k)^{-1}.
\end{equation}
The other terms correspond to SUSY breaking terms, that is, 
the soft scalar mass $m_i$ and the trilinear coupling $h^{ijk}$ 
are obtained \cite{soni,ST-soft}
\begin{eqnarray}
m_i^2 &=& m^2_{3/2} - F_a F^b\partial^a \partial_b(\ln K^i_i)+V_0,
\label{soft-m}\\
h^{ijk} &=& \hat Y^{ijk}F_a\partial^a(\tilde K - 
\ln(K^i_i K^j_jK^k_k)) +\cdots,
\label{soft-h}
\end{eqnarray}
where $\partial^a$ ($\partial_b$) denotes $\partial /\partial \Phi_a$.
Here the ellipsis denotes the term including derivatives of 
$Y^{ijk}$ by the fields $\Phi_a$.
Hereafter we restrict ourselves to the 
field-independent Yukawa couplings $Y^{ijk}$.
The other SUSY breaking terms in the scalar potential are 
suppressed by the Planck scale.
The gaugino mass $M_\alpha$ is obtained from the gauge kinetic 
function $f(\Phi_a)_\alpha$,
\begin{equation}
M_\alpha = F_a \partial^a (\ln Re f_\alpha),
\label{soft-M}
\end{equation}
and the gauge coupling $g$ is obtained 
\begin{equation}
1/g_\alpha^2 = Re(f_\alpha).
\end{equation}

The weakly coupled heterotic string model leads to 
the following gauge kinetic function and the K\"ahler potential,
\begin{eqnarray}
f_\alpha &=& S,\\
K &=& -\ln (S + \bar S) - 3 \ln (T+ \bar T) + 
3(T+ \bar T)^{n_i}|\Phi_i|^2,
\label{K-weak}
\end{eqnarray}
where $S$ and $T$ are the dilaton field and the overall moduli field, 
and $n_i$ are modular weights of $\Phi_i$.\footnote{
In general, the level of gauge group $k_\alpha$ is not 
equal to unity, but here we restrict ourselves to the case 
with $k_\alpha =1$.}

Here we restrict ourselves to the case with the universal 
K\"ahler metric $n_i=-1$, which correspond to the 
orbifold models with only untwisted matter fields and 
the large $T$ limit of Calabi-Yau models.
Because only the strongly coupled heterotic model 
with the universal K\"ahler metric has been known and 
it is not clear that we can have twisted sectors in $N=1$ 
strongly coupled models.

Here we assume that $F$-terms of $S$ and $T$ contribute 
SUSY breaking with vanishing vacuum energy, $V_0 =0$.
Following Ref.\cite{BIM}, we parameterize the $F$-terms 
by the goldstino angle $\theta$, 
\begin{eqnarray}
F^S &=& \sqrt{3} m_{3/2} (S + \bar S)\sin \theta e^{-i \gamma_S}, 
\nonumber \\
F^T &=&  m_{3/2} (T + \bar T)\cos \theta e^{-i \gamma_T},
\label{Fterms}
\end{eqnarray}
where $\gamma_S$ and $\gamma_T$ are phases of $F^S$ and $F^T$.
Hereafter we neglect CP phases $\gamma_S$ and $\gamma_T$ 
for simplicity.
Then we have 
\begin{eqnarray}
M &=& \sqrt{3} m_{3/2} \sin \theta,\\
m^2_i &=& m_{3/2}^2\sin^2 \theta,\\
h^{ijk} &=& -\sqrt{3} \hat Y^{ijk}m_{3/2} \sin \theta .
\end{eqnarray}
Thus, the relations (\ref{M-A}) and (\ref{M-m2}) hold in 
this case.
These relations are obtained even if we take into account 
the CP phases, $\gamma_S$ and $\gamma_T$.
Furthermore, we have the linear relation between the 
gauge coupling $g$ and the Yukawa coupling, 
\begin{equation}
|\hat Y^{ijk}|^2 = g^2|Y^{ijk}|^2/27.
\label{g-Y}
\end{equation}
Note that the gauge coupling is obtained $1/g^2 = Re S$.

Here we give a comment on the case with many modular weights $n_i$. 
Let us consider $T$-duality transformation,
\begin{eqnarray}
T & \rightarrow & (aT - ib)/(icT +d), \\
\Phi_i & \rightarrow & (icT + d)^{n_i} \Phi_i,
\end{eqnarray}
where $a, b, c$ and $d$ are integers satisfying 
$ad-bc =1$.
Here we require $G$ as well as the potential $V$ should be 
invariant under the above duality transformation.
That implies that the superpotential $W$ should transform 
$W \rightarrow (icT + d)^{-3}W$.
Therefore, the Yukawa couplings are allowed for the fields whose 
modular weights satisfy $n_i + n_j + n_k = - 3$.
In this case, we have the relations (\ref{M-A}),(\ref{M-m}).
Also, in this case, we have the linear relation between 
the gauge and Yukawa couplings (\ref{g-Y}).
These results are extended to the case with three diagonal moduli 
fields \cite{multiT,BIMS}.

\section{Heterotic M-theory}

Strongly coupled $E_8 \times E_8$ heterotic string theory 
has been considered as M-theory compactified on $S^1/Z_2$.
Its effective supergravity theory has been obtained 
at order of $\kappa^{4/3}$.
That is, in the strongly coupled case, $\alpha T /S $
is sizable, where $\alpha$ is written by the anomaly, and 
the gauge kinetic functions and K\"ahler potential have 
correction terms.
The corrections have been known up to $O((\alpha T /S)^2)$,
\begin{eqnarray}
f_o &=& S(1+\alpha T/S ), \qquad f_h=S(1-\alpha T/S ), 
\label{strong-f} \\
K &=& -\ln(S+\bar S) -3\ln(T + \bar T) \nonumber \\ 
&+& {3 \over T + \bar T} \left[ 1 
+ {\alpha (T + \bar T) \over 3 (S+\bar S)}\right]|\Phi_o|^2 
+{3 \over T + \bar T} \left[ 1 -
 {\alpha (T + \bar T) \over 3 (S+\bar S)}\right]|\Phi_h|^2,
\label{strong-K}
\end{eqnarray}
where $f_o$ and $f_h$ are the gauge kinetic functions of 
the observable and hidden sectors, respectively and similarly
$\Phi_o$ and $\Phi_h$ denote chiral superfields of the 
untwisted observable and hidden sectors, which correspond 
to $n_i=-1$ in eq.(\ref{K-weak}) for the weakly coupled case.
It is remarkable that the K\"ahler potential is known up to 
$O((\alpha T/S)^2)$ and thus any calculation is reliable up to 
$O((\alpha T/S)^2)$.
Only this viewpoint is different from previous calculations 
of SUSY breaking terms in heterotic M-theory.

Here we assume the following superpotential,
\begin{equation}
W = \tilde W(S,T) + {1 \over 6} Y_o^{ijk}\Phi_{oi} \Phi_{oj}\Phi_{ok}
+ {1 \over 6} Y_h^{ijk}\Phi_{hi} \Phi_{hj}\Phi_{hk},
\end{equation}
where $ \tilde W(S,T)$ is a nonperturbative superpotential 
of $S$ and $T$.
We restrict ourselves to the case that 
the Yukawa couplings $Y_o^{ijk}$ and $Y_h^{ijk}$ 
are field-independent.

Now let us assume that $F$-terms of $S$ and $T$ contribute 
SUSY breaking and the vacuum energy $V_0$ vanish.
We can use the same parameterization of $F$-terms as eq.(\ref{Fterms}) 
because the K\"ahler metric of $S$ and $T$ is same.
We use the formulae eqs.(\ref{soft-m}), (\ref{soft-h}) and 
(\ref{soft-M}) for eqs.(\ref{strong-f}) and (\ref{strong-K}).
Then we obtain the gaugino mass $M_o$, the trilinear coupling 
$h_o^{ijk}$ and the soft scalar mass $m_o$ in the 
observable sector,
\begin{eqnarray}
M_o&=&\frac{\sqrt{3}m_{3/2}}{(S+\bar{S})+\alpha(T+\bar{T})}
\left((S+\bar{S})
   \sin\theta  \right. 
+ \left. \frac{\alpha(T+\bar{T})}{\sqrt{3}}\cos\theta 
   \right) \label{Mg},\\
m^2_o&=&m^2_{3/2}-\frac{3m^2_{3/2}}{3(S+\bar{S})+\alpha(T+\bar{T})} 
     \nonumber\\
&{}& \times 
 \left\{
    \alpha(T+\bar{T})\left(2-\frac{\alpha(T+\bar{T})}{3(S+\bar{S})
       +\alpha(T+\bar{T})}\right)\sin^2\theta \right.  \nonumber\\
&{}& \left. +(S+\bar{S})\left(2-\frac{3(S+\bar{S})}{3(S+\bar{S})
       +\alpha(T+\bar{T})}\right)\cos^2\theta \right.\nonumber\\
&{}& \left. -\frac{2\sqrt{3}\alpha(T+\bar{T})(S+\bar{S})}{3(S+\bar{S})
       +\alpha(T+\bar{T})}\sin\theta\cos\theta
 \right\} \label{ms},\\
h_o^{ijk}&=&\sqrt{3}m_{3/2}\hat Y_o^{ijk}
 \left\{
    \left(-1+\frac{3\alpha(T+\bar{T})}{3(S+\bar{S})
       +\alpha(T+\bar{T})}\right)\sin\theta  \right.\nonumber\\
&{}& \left. +\sqrt{3}\left(-1+\frac{3(S+\bar{S})}{3(S+\bar{S})
       +\alpha(T+\bar{T})}\right)\cos\theta 
 \right\}. \label{A}
\end{eqnarray}
The same soft SUSY breaking terms are obtained for 
the hidden sector except replacing $\alpha$ by $-\alpha$.

Since we know  the K\"ahler 
potential up to $O((\alpha S/T)^2)$, we compare these soft 
SUSY breaking terms up to $O((\alpha S/T)^2)$, and then find 
\begin{eqnarray}
h_o^{ijk} &=& - \hat Y_o^{ijk}M_o + O((\alpha S/T)^2),\\
m_o^2 &=& {1 \over 3} M_o^2+ O((\alpha S/T)^2).
\end{eqnarray}
Note that in these relations there is no correction term 
linear to $\alpha T/S$.
The same relations  are obtained for the hidden sector, 
although the gaugino mass in the hidden sector $M_h$ 
is different from $M_o$.
If there is a matter field charged under both observable and 
hidden gauge groups (like a twisted matter field in orbifold models), 
one could see the difference between $M_o$ and $M_h$.
Similar calculations have been done in Ref.\cite{li3}, 
where the square root of 
eq.(\ref{ms}) was taken and it was deduced that the relation 
(\ref{M-m}) for 
the soft scalar masses have a correction of $O(\alpha S/T)$.
However, the soft mass squared $m^2$ is much more essential 
than the soft mass $m$ itself from the viewpoint that 
the soft mass squared $m^2$ itself is derived from the scalar 
potential (\ref{s-V}).
Thus, it is more significant to compare $m^2$ than $m$.

We have neglected the CP phases $\gamma_S$ and $\gamma_T$.
However, we obtain the same results (28) and (29) even if 
we take into account the CP phases.
Then, here we conclude that the relations (\ref{M-A}) 
and (\ref{M-m2}) still hold 
for heterotic M-theory up to $O((\alpha S/T)^2)$.
Thus, we can not distinguish soft SUSY breaking terms 
between weakly and strongly coupled heterotic string models.
Of course, the ratio of the gaugino mass to 
the gravitino mass is different between them, but 
that includes the goldstino angle, which is a free parameter. 
Also, as said above, the gaugino masses $M_o$ and $M_h$ 
in the observable and hidden sectors are different from each other, 
but this difference does not appear, 
because the observable and hidden sectors are completely 
decoupled and there is no mixing like a twisted sector of orbifold
models.
Furthermore, the linear relations between the gauge couplings 
and Yukawa couplings still hold,
\begin{equation}
(\hat Y_o^{ijk})^2 \propto g_o^2 , \qquad 
(\hat Y_h^{ijk})^2 \propto g_h^2 . 
\end{equation}
Note that the gauge couplings of the observable and hidden sectors are 
obtained $1/g_o^2 = Re(S+\alpha T)$ and $1/g_h^2 = Re(S - \alpha T)$.

All of the results, the relations of soft SUSY breaking 
terms and the linear relations 
between the gauge and Yukawa couplings would be understood if 
we write the K\"ahler metric $K_{o\ i}^i$ of the observable 
chiral field $\Phi_o$ as follows,
\begin{equation}
K_{o\ i}^i = {3 \over T + \bar T}[1+\alpha (T + \bar T)/(S + \bar S)]^{1/3} 
+O((\alpha T/S)^2).
\end{equation}
We need to know higher order corrections of K\"ahler metric 
to distinguish the weakly and strongly coupled 
heterotic string models.
If the K\"ahler metric $K_{o\ i}^i$ of the observable 
chiral field $\Phi_o$ is obtained 
\begin{equation}
K_{o\ i}^i = {3 \over T + \bar T}[1+\alpha (T + \bar T)/(S + \bar S)]^{1/3} 
+O((\alpha T/S)^n),
\end{equation}
we would have the relations (\ref{M-A}) and (\ref{M-m2}) 
up to $O((\alpha T/S)^n)$ 
and could not distinguish at this order.
In fact, this belongs to the class of supergravity models 
leading to the RG-invariant relations, 
which have been discussed in Ref. \cite{sum-rule}.

Of course, the presence of 5-branes is non-trivial in heterotic 
M-theory.
Thus, if $F$-terms of the 5-brane moduli fields contribute 
to SUSY breaking, that gives a certain type of difference from 
the weakly coupled heterotic string models as discussed in 
Ref.\cite{5brane}-\cite{BKL2}.
(However, with suppressed F-terms of 5-brane moduli fields, 
we can not see the difference.)


Here we give a comment on $\sigma$-model corrections.
Including the $\sigma$-models corrections, 
the term $-3 \ln (T + \bar T)$ of K\"ahler potential in eqs.(14) and (23) 
is shifted \cite{choi},
\begin{equation}
 -3 \ln (T + \bar T) \rightarrow  -3 \ln (T + \bar T) +{A_{03} 
\over (T + \bar T)^3},
\end{equation}
and the K\"ahler metric of $\Phi$ is corrected 
\begin{equation}
{3 \over (T + \bar T)}|\Phi|^2  \rightarrow 
{3  \over (T + \bar T)}\left(1+{B_{03} 
\over (T + \bar T)^3} \right)|\Phi|^2 ,
\end{equation}
for the weakly coupled model and 
\begin{equation}
{3 \over (T + \bar T)}\left(1+{\alpha (T + \bar T) \over 3(S+\bar S)}
\right)|\Phi_o|^2  \rightarrow 
{3  \over (T + \bar T)}\left(1+{\alpha (T + \bar T) \over 3(S+\bar S)}+
{B_{03} \over (T + \bar T)^3} \right)|\Phi_o|^2 ,
\end{equation}
for the observable sector of the strongly coupled model, 
where the coefficients $A_{03}$ and $B_{03}$ are of $O(1)$ and depend on 
geometrical aspects of the Calabi-Yau manifold \cite{sigma,CKM}.
In the orbifold case, the coefficients $A_{03}$ and $B_{03}$ 
vanish and there is no such correction.
While the parameterization of $F^S$ (15) holds true, the parameterization 
of $F^T$ changes, 
\begin{equation}
F^T = m_{3/2} (T + \bar T) \left( 1 - {2A_{03} \over (T + \bar T)^3}\right)
\cos \theta.
\end{equation}
In both of weakly and strongly coupled models, we have 
\begin{eqnarray}
h^{ijk} &=& - \hat Y^{ijk} M \left(1 + \sqrt{3} {A_{03} -3B_{03} \over 
(T + \bar T)^3}\cot \theta \right), \\ 
m^2 &=& {1 \over 3} M^2 \left( 1+  {4(A_{03} -3B_{03}) \over 
(T + \bar T)^3}\cot^2 \theta  \right),
\end{eqnarray}
up to $O(1/(T + \bar T)^6)$ and $O((\alpha T/S)^2)$.
For the model with $A_{03} = 3B_{03}$, such corrections vanish, 
but for other models, this type of corrections lead to 
significant difference between different weakly coupled Calabi-Yau
models.
However, such correction becomes less important in the 
strongly coupled model, which corresponds to 
$1/(T + \bar T)^3 << 1$.\footnote{See also Ref.\cite{CKM}.}
Therefore, in the strongly coupled model we have 
rather the universal relations (\ref{M-A}) and (\ref{M-m2}) for 
different values of $A_{03}$ and $B_{03}$.
The deviation from (\ref{M-A}), (\ref{M-m2}) would imply 
effects due to $A_{03}$ and $B_{03}$ rather in weakly coupled
models.

\section{Conclusions}

We have studied soft SUSY breaking terms 
in the strongly coupled heterotic string model.
We have found that weakly and strongly coupled models 
lead to the same relations (\ref{M-A}), (\ref{M-m2}) 
up to $O((\alpha T /S)^2)$.
Thus we can not distinguish them at the present knowledge and 
we need to know higher order of the K\"ahler potential, 
but such calculations are difficult.

\section*{Acknowledgment}

The author would like to thank K.~Choi, Y.~Kawamura and 
A.~Lukas for useful discussions.
He also thanks the Summer Institute 2000, 
Yamanashi, Japan.

\end{document}